# Superluminal Geometrodynamics of Braneworld Hyperdrive via Brane-Bulk Interaction

Anuar Alias* and Azman Jalar

Department of Applied Physics, Faculty of Science and Technology, Universiti Kebangsaan
Malaysia, Bangi, Selangor 43600, Malaysia

*anuaralias@ukm.edu.my

**Abstract.** Alcubierre's warp-drive was first to introduce superluminal travel from different approach. Instead of creating short cut through spacetime via a wormhole, geometrodynamics of simultaneously contract and expand spacetime were conjectured. However, both methods require exotic matter that violate energy condition. In this paper we employ the concept of braneworld that naturally influence spacetime geometrodynamics. Start from define and develop the spacetime metric of braneworld warp bubble, the expression for energy densities as function of relevant parameters which contribute to the braneworld approach of superluminal travel by hyperspace-drive were derive.

## 1. Introduction

The astronomical scale of the universe indicates that the speed of light is apparently slow even though it is considered as the speed limit of the universe. A spacefaring civilization without superluminal travel capability is so limited toward travelling within a solar system. Even though such civilization may achieve subluminal capability at 99% the speed of light, the limitation of travel is still just within a very small sector of a galaxy. Thus, the only meaningful spacefaring civilization is a civilization that has the capability to cover a travelling area at least the size of a galaxy where requirement of superluminal travel capability is necessary. Therefore motivationally, a superluminal capability that is the hyper-fast travel at many folds the speed of light is intriguing topic to study and explore which was firstly introduced by Alcubierre [1]. Superluminal travel seems to violate Special Relativity (SR), However in General Relativity (GR), it has been shown that the geometrodynamics [2] of spacetime does not constrain by any velocity limitation during the spacetime expansion or spacetime contraction. On the cosmological scale, the observable phenomenon of the universe expansion has proved this notion especially when the signature of the inflationary period [3] of the universe evolution is taken into consideration. Therefore, within the scope of GR, superluminal travel is naturally plausible. Concept of superluminal travel can be categorized into two, which are "tunnelling-through" and "surfing-on" the spacetime. The distance travelled could be reduced significantly by the tunnelling concept as in wormhole which indicate the topological breaking of spacetime. This could happen either naturally as suggested in quantum foam [4] or could be induced through entanglement [5]. A warp bubble of spacetime that is embedded in the surrounding spacetime is analogous to the concept of surfing. The hypersurface of warped bubble is "surfing-on" its surrounding hyperspace.

In this paper, we study the braneworld cosmological concept that can possibly influence the embedding mechanism of warp bubble. We will introduce a new method of warp drive using brane-bulk fluid [6] interaction that enhances the warp drive concept toward the hyperdrive concept. Braneworld [7] cosmology suggests that there exists 1-dimensional higher spatial extra-dimension than the 3-dimensional space, known as bulk, that may contain fluid-like substance described by Weyl tensor [8]. The 3-d universe is the 3-dimensional spatial brane floating in 4-dimensional spatial bulk. Thus, in deriving the warp bubble expression, the extrinsic curvature not only require to be dealt with warp bubble hypersurface as brane but also with bulk-brane interaction [9] in line with the idea that spacetime inflation is driven by geometrodynamics of the interaction.

## 2. Warp bubble's spacetime metric.

The spacetime metric that represent a warp bubble is considered to have brane-bulk interaction characteristic of $Y(r)$ as shown below

$$ds^2 = -dt^2 + \left(1+\left(\frac{dY}{dr}\right)^2\right)dr^2 + r^2 d\theta^2 + r^2 \sin^2\theta d\phi^2 . \tag{1}$$

The $Y$ term indicates brane-bulk interaction of the spacetime [8],[9]. Thus, equation (1) represents the brane spacetime that is embedded in a bulk space. The warp bubble interior is a flat spacetime where $Y \to 0$. Consider the warp-bubble move at a velocity $v$ along the $x$ axis represented as $v = v|_x$, the metric of the warp bubble that moves along the $x$-axis can be rewritten from equation (1) as

$$ds^2 = -dt^2 + \left(dr|_x - v|_x dt\right)^2 + \left(\frac{dY}{dr}\right)^2 dr^2 + r^2 d\theta^2 + r^2 \sin^2\theta d\phi^2 . \tag{2}$$

The essence of equation (1) is conserved, as the equation (2) will become (1) again if the warp bubble is at stationary condition with respect to the surrounding space. The warp bubble has range of radius from $r = 0$ at the center of the bubble to $r = R$ at the edge of the bubble.

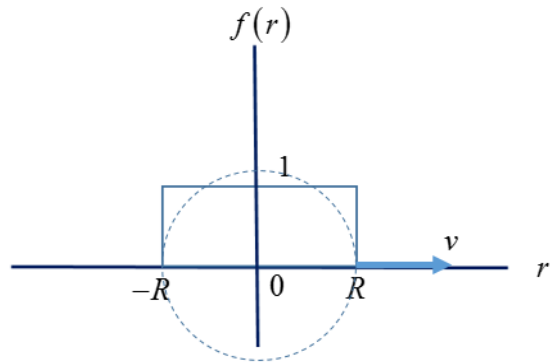

**Figure 1.** Warp bubble spacetime radii characteristic

Within the region $0 \leq |r| \leq R$, the shape function $f(r) = 1$. This characteristic is depicted in Fig. 1 below. Preserving the essence of equation (2) the shape function $f(r)$ is coupled with velocity indicating that the $v|_x f$ term correspond only to the region of the warp bubble, therefore $v$ is representing warp bubble velocity along the $x$- axis relative to its surrounding brane spacetime. The radial infinitesimal distance in $x$-axis direction is equal to general radial infinitesimal that is $dr|_x = dr$. The equation (2) can generally be written with the inclusion of the shape function as below

$$ds^2 = -dt^2 + \left(dr - v|_x f(r) dt\right)^2 + \left(\frac{dY}{dr}\right)^2 dr^2 + r^2 d\theta^2 + r^2 \sin^2\theta d\phi^2 . \tag{3}$$

It is obvious that the spatial coordinate at the center of the bubble is $(r|_x, \theta, \phi) = (v|_x ft, 0, 0)$ Therefore, the warp bubble spacetime metric reduced to $ds^2 = -dt^2$ since $dr|_x = v|_x fdt$ and $d\theta = d\phi = 0$. Imposing $dt^2 = -ds^2 = d\tau^2$, infers the characteristics of the bubble's *null-time dilation* where the proper time is equal to the $\tau = t$. This indicates that the spacetime condition inside the bubble is equivalent to a stationary condition that is non-relativistic. Expanding this notion, we may obtain geodesic characteristic of the warp bubble. Thus by $d\tau^2 = -ds^2 = -g_{\mu\nu}dx^{\mu}dx^{\nu}$, the proper time can be written as below to consist of the Lagrangian terms that represent curvature invariants of the warp bubble [10]

$$\tau = \int \sqrt{-g_{\mu\nu}\dot{x}^{\mu}\dot{x}^{\nu}}\,d\lambda, \tag{4}$$

This is the proper time's extremum [1] integral expression of the Lagrangian term with respect to a parameter $\lambda$ (which later to be assigned as proper time $\tau$). The Lagrangian can be obtained by expand and rearrange a flat spacetime metric equation inside the warp bubble. Thus, the extradimensional term $Y$ that represents the bulk space is negligible from the perspective of a Eulerian observer inside the bubble. Therefore, as $Y \simeq 0$ we may write

$$ds^2 = -dt^2 + \left(dr - v|_x f(r)dt\right)^2 + r^2 d\theta^2 + r^2 \sin^2\theta d\phi^2,$$
$$= -\left(1 - v|_x^2 f(r)^2\right)dt^2 - 2v|_x f(r)drdt + dr^2 + r^2 d\theta^2 + r^2 \sin^2\theta d\phi^2, \tag{5}$$

so, the following are the metric components identified.

$$g_{00} = v|_x^2 f(r)^2 - 1,\ g_{01} = g_{10} = -v|_x f(r),\ g_{11} = 1,\ g_{22} = r^2,\ g_{33} = r^2 \sin^2\theta, \tag{6}$$

by (4) and let generalizing $v|_x = v$

$$\left(\frac{d\tau}{d\lambda}\right)^2 = \frac{d\tau^2}{d\lambda^2} = \left(1 - v^2 f^2\right)\dot{t}^2 + 2vf\dot{r}\dot{t} - \dot{r}^2 - r^2\dot{\theta}^2 - r^2\sin^2\theta\dot{\phi}^2 = -g_{\mu\nu}\dot{x}^{\mu}\dot{x}^{\nu}. \tag{7}$$

From the pure definition of Lagrangian representing a kinetic energy; $L = \frac{1}{2}\dot{q}^2$ where $q = s(x^{\mu})$ as a coordinate position of a metric where $x^{\mu} = x^{\mu}(t)$, thus

$$\dot{q}^2 = \dot{s}^2 = (-\dot{\tau})^2 = \frac{-d\tau^2}{d\lambda^2} = \frac{ds^2}{d\lambda^2} = \left(\frac{-d\tau}{d\lambda}\right)^2 = \left(\frac{ds}{d\lambda}\right)^2 = 2L \tag{8}$$

since $-ds^2 = d\tau^2$; $ds^2 = -d\tau^2$ and the dot indicates the derivative with respect to $\lambda$.

From (7) and (8) $-g_{\mu\nu}\dot{x}^{\mu}\dot{x}^{\nu} = \left(1 - v^2 f^2\right)\dot{t}^2 + 2vf\dot{r}\dot{t} - \dot{r}^2 - r^2\dot{\theta}^2 - r^2\sin^2\theta\dot{\phi}^2 = \frac{d\tau^2}{d\lambda^2} = \frac{-ds^2}{d\lambda^2}$ therefore

$$\left(\frac{ds}{d\lambda}\right)^2 = \frac{ds^2}{d\lambda^2} = \frac{-d\tau^2}{d\lambda^2} = \dot{q}^2 = g_{\mu\nu}\dot{x}^{\mu}\dot{x}^{\nu} = -\left(1 - v^2 f^2\right)\dot{t}^2 - 2vf\dot{r}\dot{t} + \dot{r}^2 + r^2\dot{\theta}^2 + r^2\sin^2\theta\dot{\phi}^2 = 2L,$$

therefore

$$L = \frac{1}{2}\left(-\left(1 - v^2 f^2\right)\dot{t}^2 - 2vf\dot{r}\dot{t} + \dot{r}^2 + r^2\dot{\theta}^2 + r^2\sin^2\theta\dot{\phi}^2\right). \tag{9}$$

Rearrange equation (9) we may have the Lagrangian that represents the warp bubble motion,

$$L = \frac{1}{2}\left(-\dot{t}^2 + r^2\dot{\theta}^2 + r^2\sin^2\theta\dot{\phi}^2 + \left(\dot{r} - vf\dot{t}\right)^2\right). \tag{10}$$

Parameterizing the geodesic using proper time by taking $\lambda = \tau$ and operate (10) by using Euler Lagrange $\frac{\partial}{\partial \tau}\left(\frac{\partial L}{\partial \dot{x}^\mu}\right) - \frac{\partial L}{\partial x^\mu} = 0$. Taking $\partial \dot{x}^\mu = \frac{\partial x^\mu}{\partial \tau}$, we may write as follows,

for $\mu = t$

$$\frac{\partial}{\partial \tau}\left(-\dot{t} - vf\left(\dot{r} - vf\dot{t}\right)\right) + \left(\dot{r} - vf\dot{t}\right)\dot{t}\frac{d(vf)}{dt} = 0 \ , \tag{11}$$

for $\mu = r$
$$\frac{\partial}{\partial \tau}\left(\dot{r} - vf\dot{t}\right) + \left(\dot{r} - vf\dot{t}\right)\dot{t}\frac{\partial(vf)}{\partial r} = 0 \ , \tag{12}$$

for $\mu = \theta$
$$\frac{\partial \dot{\theta}}{\partial \tau} + \left(\dot{r} - vf\dot{t}\right)\dot{t}\frac{\partial(vf)}{\partial \theta} = 0 \ , \tag{13}$$

and for $\mu = \phi$
$$\frac{\partial \dot{\phi}}{\partial \tau} + \left(\dot{r} - vf\dot{t}\right)\dot{t}\frac{\partial(vf)}{\partial \phi} = 0 \ . \tag{14}$$

Equations (11) to (14) indicate that the solution shall be

$$\dot{x}^\mu = \left(\dot{x}^0, \dot{x}^1, \dot{x}^2, \dot{x}^3\right) = \left(\dot{t}, \dot{r}, \dot{\theta}, \dot{\phi}\right) = \left(1, vf\dot{t}, 0, 0\right). \tag{15}$$

These are 4 velocities on timelike geodesics of Eulerian observers. The velocity represented in terms of shape function $\dot{x} = vf\dot{t}$. At $r = 0$, that is at the middle of the bubble interior, the shape function is valued as $f = 1$ and the parameter $\lambda = \tau$ a proper time, thus as $\dot{t} = dt/d\lambda$, therefore $\dot{t} = 1$ due to $t = \tau$ in the interior of the warp bubble. Since there is no warp bubble's corresponding mass or length constraints its velocity $v$ in equation (15) is not limited to the speed of light $c$. Using the metric equations (6), the interior of warp bubble can also be shown to possess timelike characteristic. By contracting with the metric tensor, we may extract the derivative of $\dot{x}_\mu = g_{\mu\nu}\dot{x}^\nu$.

For $\dot{x}_0 = g_{0\nu}\dot{x}^\nu$, since $g_{02} = g_{03} = 0$, $\frac{\partial t}{\partial \tau} = 1$, $\frac{\partial x}{\partial \tau} = \frac{\partial x}{\partial t} = v$ and $f = 1$

$$\dot{x}_0 = \left(v^2 f^2 - 1\right)\frac{\partial t}{\partial \tau} + \left(-vf\right)\frac{\partial x}{\partial \tau} = -1 \ . \tag{16}$$

For $\dot{x}_1 = g_{1\nu}\dot{x}^\nu$, since $g_{12} = g_{13} = 0$, $\frac{\partial t}{\partial \tau} = 1$, $\frac{\partial x}{\partial \tau} = \frac{\partial x}{\partial t} = v$ and $f = 1$

$$\dot{x}_1 = \left(-vf\right)\frac{\partial t}{\partial \tau} + \frac{\partial x}{\partial \tau} = 0. \tag{17}$$

The rest, that is $\dot{x}_2 = \dot{x}_3 = 0$ obviously, which therefore

$$\dot{x}_\mu = \left(-1, 0, 0, 0\right) \tag{18}$$

The derivatives are normal to the hypersurface [1] as $dt = 0$ and by equation (15) $\dot{x}_\mu \dot{x}^\mu = -1$ confirming the warp bubble timelike characteristic. Thus, ensure that any object inside the bubble can be at stationary or moving at sub-light speed that does not violate special relativity.

Let's define $\beta = -vf(r)$ so we may write equation (5) as follows

$$ds^2 = -(1-\beta^2)dt^2 + 2\beta dr dt + dr^2 + r^2 d\theta^2 + r^2 \sin^2\theta d\phi^2 . \tag{19}$$

All the terms in equation (19) can be represented in tensorial form as the following

$$\beta^2 = \beta^i \beta_i , \quad dr^2 + r^2 d\theta^2 + r^2 \sin^2\theta d\phi^2 = \delta_{ij} dx^i dx^j \text{ and } \beta dx = \beta_i dx^i \tag{20}$$

where $i,j = 1,2,3$, implies that $v = -\beta$ is in the bubble's direction of travel, otherwise $v = 0$. In this case we consider warp bubble analogously surfs along the $x$ axis, therefore $\beta_1 = \beta_{r|x} = -v$ and $\beta_2, \beta_3 = \beta_{r|y}, \beta_{r|z} = 0$. Using (20), in a generalized tensor form, equation (19) becomes

$$ds^2 = -(1 - \beta^i \beta_i) dt^2 + 2\beta_i dx^i dt + \delta_{ij} dx^i dx^j . \tag{21}$$

The term $\alpha = \sqrt{\alpha^i \alpha_i} = 1$ is introduced to generalize the metric tensor $g_{00}$ of equation (21), thus we can show that

$$ds^2 = -(\alpha^2 - \beta^i \beta_i) dt^2 + 2\beta_i dx^i dt + \delta_{ij} dx^i dx^j \tag{22}$$

as firstly introduced by Alqubierre [1]. All these imply that a warp bubble spacetime metric that has been considered to have a brane-bulk interaction characteristic, can be deduced to the original Alqubierre metric. Therefore, from this notion, we may workout in similar manner for brane-bulk space drive.

## 3. Shape function

The shape function is a function of the warp bubble radius $f = f(r)$. The radius $r$ is measured from the center of the bubble's interior. In Figure 2 it is constructed as $f(r)|_{|r|<R} = 1$ for the bubble's interior and $f(r)|_{|r|>R} = 0$ for the bubble's exterior. It can be shown [1] that in general, the function can be written as

$$f(r) = \frac{\tanh(\sigma(r+R)) - \tanh(\sigma(r-R))}{2\tanh(\sigma R)}, \tag{23}$$

where $\sigma$ is a factor that corresponds to the thickness of the bubble. The details of the shape function character can be described by the following four cases with respect to the warp bubble's radii:

*3.1 Positive radius of the bubble along the direction of travel ; $r = R$.*

The transition of shape function from $f = 1$ to $f = 0$, occurs at the positive side of the radius where $r \to R$ is approaching the bubble's wall toward the direction of travel. Instead of an abrupt transition, equation (23) ensures a gradual transition character. The contraction of spacetime occurs in the transition region at the front of the bubble along the direction of travel.

At exactly $r = R$, the shape function, equation (23) is deduced by $f(r)\big|_{r=R} = \dfrac{\tanh(\sigma(2R))}{2\tanh(\sigma R)}$, which then becomes $\dfrac{\tanh(\sigma(R)) + \tanh(\sigma(R))}{2\tanh(\sigma R)(1 + \tanh^2(\sigma(R)))}$, thus

$$f(r)\big|_{r=R} = \frac{1}{1 + \tanh^2(\sigma R)} \quad . \tag{24}$$

This implies that the bubble's thickness factor $\sigma$ influence the transition characteristic. As $\sigma \to 0$, the transition from $f(r)\big|_{|r|<R} = 1$ toward $f(r)\big|_{|r|>R} = 0$ is abrupt.

*3.2 Warp bubble's central position; $r = 0$*

The bubble interior region as $r \to 0$ where objects in this region experience flat spacetime. As $r = 0$, the shape function, equation (23) is deduced to 1 by $f(r)\big|_{r=0} = \dfrac{\tanh(\sigma(R)) - \tanh(\sigma(-R))}{2\tanh(\sigma R)}$ and since $\tanh(\sigma(-R)) = -\tanh(\sigma(R))$,

$$f(r)\big|_{r=0} = \frac{\tanh(\sigma(R)) + \tanh(\sigma(R))}{2\tanh(\sigma R)} = 1 \quad . \tag{25}$$

*3.3 Negative radius of the bubble opposing the direction of travel; $r = -R$*

Similar to the case in *3.1* the negative side of the bubble radius where $r \to -R$ which is approaching the bubble's wall toward the opposite direction of travel, the transition of shape function from $f = 1$ to $f = 0$ occurs. The equation (23) ensures the transition graduality instead of abruptness. As the bubble moves in the direction of travel where the spacetime at the frontal region contract (case 3.1), the transition region in this case represents the spacetime expansion at the rear of the warp bubble. At $r = -R$, equation (23) deduced as in the case of *3.1*, where $f(r)\big|_{r=-R} = \dfrac{-\tanh(\sigma(-2R))}{2\tanh(\sigma R)}$ and since $\tanh(\sigma(-2R)) = -\tanh(\sigma(2R))$, thus $\dfrac{\tanh(\sigma(2R))}{2\tanh(\sigma R)} = \dfrac{\tanh(\sigma(R)) + \tanh(\sigma(R))}{2\tanh(\sigma R)(1 + \tanh^2(\sigma(R)))}$, therefore

$$f(r)\big|_{r=-R} = \frac{1}{1 + \tanh^2(\sigma R)} \quad . \tag{26}$$

*3.4 The exterior region of the bubble; $|r| > R$*

The spacetime of the bubble's exterior region exhibit flatness characteristic. Along the direction of travel, the shape function shows gradual transitions of $f = 0 \to f = 1$. This indicates that the spacetime expand at the rear of the bubble. At the front of the bubble, the shape function shows gradual transitions of $f = 1 \to f = 0$ indicating spacetime contraction. By equation (23), it can be shown that as $r < -R$, $(r|_{r<-R} + R) < 0$ and $(r|_{r<-R} - R) < -2R$ therefore

$$f(r)\big|_{r<0} = \frac{\tanh\left(\sigma\left(\left(r|_{r<-R}+R\right)<0\right)\right) - \tanh\left(\sigma\left(\left(r|_{r<-R}-R\right)<-2R\right)\right)}{2\tanh(\sigma R)},$$

since $\tanh\left(\sigma\left(\left(r|_{r<-R}+R\right)<0\right)\right) \approx -1$ and $\tanh\left(\sigma\left(\left(r|_{r<-R}-R\right)<-2R\right)\right) \approx -1$ thus $f(r)\big|_{r<-R} \approx 0$

as $r > R$, $\left(r|_{r>R}+R\right) > 2R$ and $\left(r|_{r>R}-R\right) > 0$ therefore

$$f(r)\big|_{r>R} = \frac{\tanh\left(\sigma\left(\left(r|_{r>R}+R\right)>2R\right)\right) - \tanh\left(\sigma\left(\left(r|_{r>R}-R\right)>0\right)\right)}{2\tanh(\sigma R)},$$

since $\tanh\left(\sigma\left(\left(r|_{r>R}+R\right)>2R\right)\right) \approx 1$ and $\tanh\left(\sigma\left(r|_{r>R}-R\right)>0\right) \approx 1$ thus $f(r)\big|_{r>R} \approx 0$

thus

$$f(r)\big|_{|r|>R} \approx 0 \quad . \tag{27}$$

Equations (24), (25), (26) and (27), have shown that the bubble's thickness factor $\sigma$ in the equation (23) not only ensure that the shape function transition in Figure 2 becoming gradual but also indicate the spacetime contraction and expansion dynamics along the bubble's direction of travel.

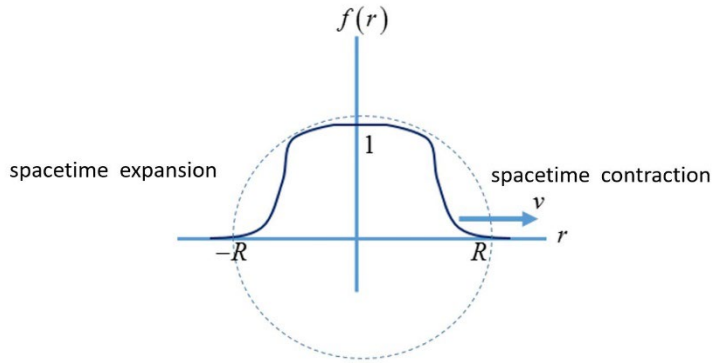

**Figure 2.** Bubble radii with gradual transition characteristic

The transition region is vital for the warp bubble dynamics. The parameters that defined the warp bubble character are the governing factors for the dynamic characteristic.

**4. Spacetime expansion and contraction dynamics**
The extrinsic curvature relationship between the warp bubble and the surrounding spacetime is shown in figure 3.

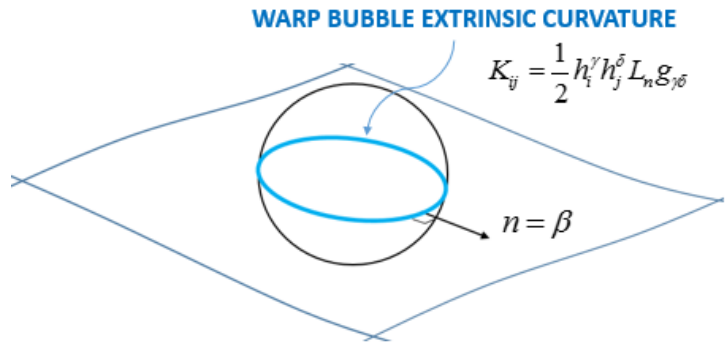

**Figure 3.** Warp bubble and the surrounding spacetime intrinsic curvature

Let the spacetime of warp bubble and its surrounding defined as "locality". The metric that represents the locality is a combination of two flat spacetime which are the warp bubble with flat spacetime interior embedded into the surrounding flat spacetime. The dynamics of embedding a warped spacetime hypersurface onto the surrounding hyperspace is analogous to *spacetime surfing*. The transition region where the surrounding hyperspace interact with the bubble is represented by the extrinsic curvature. The extrinsic curvature is shown by the Lie derivative of the projection tensor $h_{ij} = h_i^\gamma h_j^\delta g_{\gamma\delta}$. The derivative indicates the projection tensor's rate of change that is projected along the normal vector field $n_i$ [7]. Physically, the vector field is normal to the warped spacetime hypersurface of the warp-bubble.

$$K_{ij} = \frac{1}{2} h_i^\gamma h_j^\delta L_n g_{\gamma\delta}. \tag{28}$$

The normal vector is representing the direction of travel of the warp bubble $n = \sqrt{n_i n^i} = \sqrt{\beta_i \beta^i} = \beta$. Thus, we may rewrite (28) as follows

$$K_{ij} = \frac{1}{2} h_i^\gamma h_j^\delta L_\beta g_{\gamma\delta}, \tag{29}$$

which then expanded as the warp bubble covariant derivatives in the direction of travel-vector $\beta_i$. Equation (29) reduces toward the expression derived elaborately as the following

$$\begin{aligned} K_{ij} &= \frac{1}{2} h_i^\gamma h_j^\delta \left( \nabla_\gamma \beta_\delta + \nabla_\delta \beta_\gamma \right) \\ &= \frac{1}{2} h_i^\gamma h_j^\delta \left( \partial_\gamma \beta_\delta - \Gamma^\lambda_{\gamma\delta} \beta_\lambda + \partial_\delta \beta_\gamma - \Gamma^\lambda_{\delta\gamma} \beta_\lambda \right) \\ &= \frac{1}{2} \left( \partial_i \beta_j + \partial_j \beta_i + \beta^0 \partial_0 g_{ij} + \beta^1 \partial_1 g_{ij} + \beta^2 \partial_2 g_{ij} + \beta^3 \partial_3 g_{ij} \right) \end{aligned} \tag{30}$$

Considering flat spacetime surrounding the warp bubble, thus approximately $\beta^i \partial_i g_{ij} \simeq 0$, since $g_{ij} = g_{xx} = g_{yy} = g_{zz} = 1$, therefore the equation (30) reduced to

$$K_{ij} = \frac{1}{2} \left( \partial_i \beta_j + \partial_j \beta_i \right), \tag{31}$$

which is the warp bubble equation of motion. The extrinsic curvature in matrix form can be written as

$$K = \begin{pmatrix} K_{11} & K_{12} & K_{13} \\ K_{21} & K_{22} & K_{23} \\ K_{31} & K_{32} & K_{33} \end{pmatrix}, \tag{32}$$

The trace of the matrix indicates the dynamic's characteristic $\chi$ of space surrounding the warp bubble of either expansion or contraction

$$\begin{aligned} \chi &= -\alpha TrK, \\ &= -\alpha \left( K_{11} + K_{22} + K_{33} \right), \\ &= -\alpha \left( \partial_r \beta_r - vf(r) \frac{\partial^2 Y}{\partial r^2} \frac{\partial Y}{\partial r} + \partial_\theta \beta_\theta + \partial_\phi \beta_\phi \right) \end{aligned} \tag{33}$$

but since $\beta_\theta = \beta_\phi = 0$,

thus, the dynamics characteristic term is reduced to

$$\chi = -\alpha\left(\partial_r \beta_r - v\frac{\partial^2 Y}{\partial r^2}\frac{\partial Y}{\partial r}\right).$$

(34)

where $v\frac{\partial^2 Y}{\partial r^2}\frac{\partial Y}{\partial r} = v_{bulk}$, is the velocity of the bulk. Therefore, the dynamics characteristic can also be expressed as $\chi = \alpha(v_{bulk} - \partial_r \beta_r)$. Beside indicating the warp bubble surrounding space of either expansion or contraction, the dynamics characteristic also indicates the bubble's proportionality with velocity. Expanding $\partial_r \beta_r$

$$\partial_r \beta_r = -\frac{\partial v f(r)}{\partial r}$$
$$= -\left(v(t)\frac{\partial f(r)}{\partial x} + f(r)\frac{\partial v(t)}{\partial x}\right) = -v(t)\frac{\partial f(r)}{\partial r} \quad (35)$$

since

$$v = v(t) = \frac{dx_b(t)}{dt}, \text{ thus } \frac{\partial v(t)}{\partial x} = \frac{\partial(dx_b/dt)}{\partial x} = 0 \quad (36)$$

It represents the velocity of the warped bubble with respect to its surrounding spacetime (in the brane) without considering the bulk space factor. Thus, it can also be considered as the on-brane velocity $v = v_{brane}$. In equation (36) and the following equations, $x_b$ is the position of the moving warp bubble with respect to its point of origin depicted in figure 4.

$$r = r(x,t) = \left[(x - x_b(t))^2 + y_b^2 + z_b^2\right]^{\frac{1}{2}}, \quad x_b = x_b(t) \text{ thus } dx_b = \frac{dx_b(t)}{dt}dt, \quad x_b = \frac{dx_b(t)}{dt}t \quad (37)$$

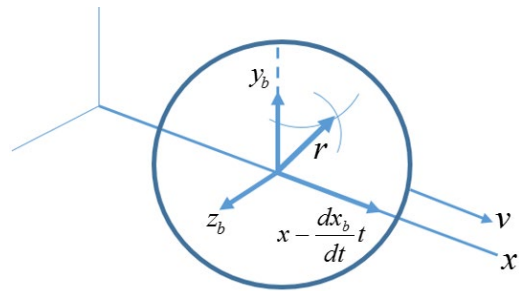

**Figure 4.** Dynamics parameters of the bubble

From (35) and (37), since $\partial_{r|x}\beta_{r|x} = -v(t)\frac{\partial f(r)}{\partial x} = -v_{brane}\frac{\partial f(r)}{\partial r}\frac{\partial r(x,t)}{\partial x}$, considering $v = v_{brane}$, it can be shown that $\partial_{r|x}\beta_{r|x} = -v_{brane}\frac{(x - x_b(t))}{r}\frac{\partial f(r)}{\partial r}$. Thus, from equation (34), the spacetime dynamic characteristic

$$\chi = -\alpha\left(\partial_{r|x}\beta_{r|x} - v_{bulk}\right) = \alpha\left(v_{brane}\frac{(x-x_b(t))}{r}\frac{\partial f(r)}{\partial r} + v_{bulk}\right). \tag{38}$$

let $\alpha = 1$ for simplicity and hyperdrive velocity is defined as $v_{HD} = v_{brane} + v_{bulk}$

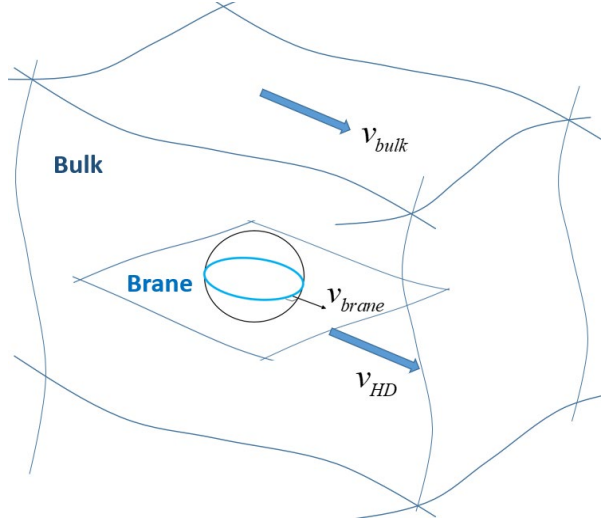

**Figure 5.** Extrinsic curvature between hyperdrive warp bubble and the surrounding spacetime

The hyperdrive velocity is an accumulation of warped bubble on-brane velocity pushed by the underlying bulk velocity as depicted in Figure 5. Analogically as if wind that represent bulk, push the kite surfing surfer that represents the warp bubble. Thus, from equation (38), the overall dynamics characteristics indicator expression is $\chi = (v_{HD} - v_{bulk})\frac{(x-x_b(t))}{r}\frac{\partial f(r)}{\partial r} + v_{bulk}$. The extrinsic curvature of the hyperdrive warped bubble with brane and bulk interaction is also depicted in Figure 6.

From this expression we may study the warp bubble regional characteristics that consist of *interior-exterior* and *transition* regions. In the *interior region* of the warp bubble, $f = 1$, while in *the exterior region* of the warp bubble, $f = 0$. Consider only the dynamics of warp bubble with respect to the surrounding space that is the brane only, thus $v_{bulk} = 0$. The warp bubble interior and exterior region spacetime dynamics characteristics can be depicted in figure 6 as

$$\chi = v_{brane}\frac{(x-x_b(t))}{r}\left(\frac{\partial f(r)}{\partial r}\right)\bigg|_{f(r<-R)=0} = 0 \qquad \chi = v_{brane}\frac{(x-x_b(t))}{r}\left(\frac{\partial f(r)}{\partial r}\right)\bigg|_{f(r>R)=0} = 0$$

**Exterior region**   $-R$   $0$   $R$   **Exterior region**

**Interior region**

$$\chi = v_{brane}\frac{(x-x_b(t))}{r}\left(\frac{\partial f(r)}{\partial r}\right)\bigg|_{f(-R<r<R)=1} = 0$$

**Figure 6**. Spacetime dynamic characteristics in the warp bubble interior and exterior regions

Thus, the expression that represent interior-exterior region is

$$\chi = v_{brane} \frac{(x - x_b(t))}{r} \left( \frac{\partial f(r)}{\partial r} \right)\Bigg|_{f(-R<r<R)=1} = v_{brane} \frac{(x - x_b(t))}{r} \left( \frac{\partial f(r)}{\partial r} \right)\Bigg|_{f(|r|>R)=0} = 0 \ . \tag{39}$$

This indicates the disconnection between warp bubble regions of interior and exterior while preserving flat non-dynamics character.

At the *transition* where $f(r): 0 \Leftrightarrow 1$, $\theta \neq 0$, as shown on figure 7, these regions constitute two thickening walls with thickness $\xi = 1/\sigma$ along the direction of the warp bubble travel.

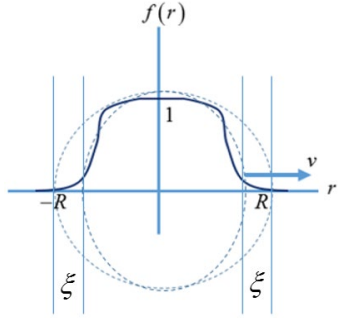

**Figure 7**. Spacetime dynamic's characteristics at the warp bubble's transition regions

The transition region in figure 8 and 9 will show the characteristics of spacetime expansion and contraction respectively. At the front of the bubble, in the direction of travel

$$\chi = v_{brane} \frac{(x - x_b(t))}{r} \left( \frac{\partial f(r)}{\partial r} \right)\Bigg|_{(f=1) \to (f=0)} < 0 \ , \tag{40}$$

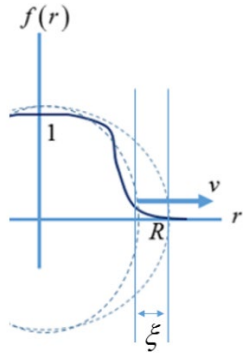

**Figure 8**. Transition region at the front of the bubble

since, $\left.\dfrac{\partial f(r)}{\partial r}\right|_{(f=1)\to(f=0)} < 0$, thus $x - x_b(t) > 0$, therefore $x_b(t) < x$, which represents *space contraction*. In the transition region at the rear of the bubble, opposite to the direction of travel (as opposed to contraction condition ($\theta < 0$)), thus $\theta > 0$ that is equation (38) is positive in value.

$$\chi = v_{brane} \dfrac{(x - x_b(t))}{r}\left(\left.\dfrac{\partial f(r)}{\partial(r)}\right|_{(f=1)\to(f=0)}\right) > 0, \qquad (41)$$

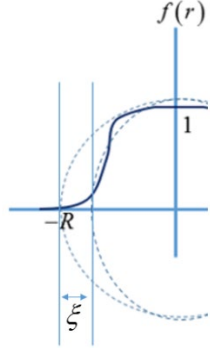

**Figure 9**. Transition region at the rear of the bubble

since also that $\left.\dfrac{\partial f(r)}{\partial(r)} = \dfrac{\partial f(-r)}{\partial(-r)}\right|_{(f=1)\to(f=0)} < 0$, which implies that $\dfrac{\partial f}{\partial r} < 0$ as always (of either during contraction or expansion), and $\theta > 0$ (as opposed to contraction condition), thus $x - x_b(t) < 0$, therefore $x_b(t) > x$, which represents *space expansion*.

## 5. Velocity expression

The relationship between the warp bubble velocity and the energy density can be acquired by expanding the Einstein curvature tensor $G_{\mu\nu}$ in terms of Ricci tensor. Rearranging the expression in terms of the energy momentum tensor and multiplying with timelike vector $t^\mu$ will result to the following

$$T_{\mu\nu}t^\mu t^\nu = \dfrac{1}{8\pi}\left(R_{\mu\nu} - \dfrac{1}{2}Rg_{\mu\nu}\right)t^\mu t^\nu. \qquad (42)$$

As proven by Lobo et. al [7] and by considering the warp bubble's direction of travel is in the $x$ direction (Fig. 5) the equation (42) becomes

$$T_{\mu\nu}t^\mu t^\nu = -\dfrac{v_{HD}^2}{32\pi}\left(\left(\dfrac{\partial f}{\partial y}\right)^2 + \left(\dfrac{\partial f}{\partial z}\right)^2\right). \qquad (43)$$

which is the energy density of the warp bubble. Since it violates the energy condition where $T_{\mu\nu}t^\mu t^\nu < 0$ thus we may define

$$T_{\mu\nu}t^\mu t^\nu = -\rho_{H\_warp}. \qquad (44)$$

The velocity term emerges in equation (43) from Lobo and Visser's calculation [17] is regarded as the overall velocity of the hyperdrive warp bubble $v_{HD}$. It is known that the shape function $f = f(r)$, thus $\partial f = \frac{\partial f(r)}{\partial r} \partial(r)$, therefore $\frac{\partial f}{\partial y} = \frac{\partial f(r)}{\partial r} \frac{\partial r}{\partial y}$ and $\frac{\partial f}{\partial z} = \frac{\partial f(r)}{\partial r} \frac{\partial r}{\partial z}$ from (37) as $y_b = y$ and $z_b = z$

$$\frac{\partial f}{\partial y} = \frac{\partial f(r)}{\partial r} \frac{\partial \left[(x-x_b(t))^2 + y^2 + z^2\right]^{\frac{1}{2}}}{\partial \left((x-x_b(t))^2 + y^2 + z^2\right)} \frac{\partial (x-x_b(t))^2 + y^2 + z^2}{\partial y} = \frac{y}{r} \frac{\partial f(r)}{\partial r} \quad (45)$$

$$\frac{\partial f}{\partial z} = \frac{\partial f(r)}{\partial r} \frac{\partial \left[(x-x_b(t))^2 + y^2 + z^2\right]^{\frac{1}{2}}}{\partial \left((x-x_b(t))^2 + y^2 + z^2\right)} \frac{\partial (x-x_b(t))^2 + y^2 + z^2}{\partial z} = \frac{z}{r} \frac{\partial f(r)}{\partial r} \quad (46)$$

by (43) to (46) the expression of the warp bubble energy density is

$$\rho_{H\_warp} = \frac{v_{HD}^2}{32\pi} \left(\frac{y^2+z^2}{r^2}\right) \left(\frac{\partial f(r)}{\partial r}\right)^2 . \quad (47)$$

The derivative $\partial f(r)/\partial r \approx \sigma$ which is actually the slope of the transition region in the condition where $f(r) : 0 \Leftrightarrow 1$ and $\theta \neq 0$ which is also the thickness factor. The thickness of the bubble is $\xi = 1/\sigma$ (Fig. 7) thus from (47) the warp bubble velocity expression can be written as

$$v_{HD} = v_{brane} + v_{bulk} \approx \xi r \sqrt{\frac{32\pi \rho_{warp}}{y^2+z^2}} = \frac{\xi r}{A} \left(32\pi \rho_{H\_warp}\right)^{\frac{1}{2}} . \quad (48)$$

This equation shows that the thickness of the warp bubble $\xi$ and the hyperdrive energy density $\rho_{H\_warp}$ are the two most important factors that determine the hyperdrive warp bubble speed $v_{HD}$. The hyperdrive energy density consists of brane density and bulk density terms as

$$\rho_{H\_warp} = \rho_{brane} + \rho_{bulk} . \quad (49)$$

From equations (47), (48) and (49) the expression of the energy densities can be shown as

$$\rho_{brane} = \frac{v^2}{32\pi} \left(\frac{A}{r} \frac{\partial f(r)}{\partial r}\right)^2 , \quad (50)$$

which is the brane energy density representing energy density that is required for the warp bubble dynamics with respect to the surrounding space, and

$$\rho_{bulk} = \frac{\left(v_{bulk}^2 + 2vv_{bulk}\right)}{32\pi} \left(\frac{A}{r} \frac{\partial f(r)}{\partial r}\right)^2 , \quad (51)$$

which is the bulk energy density representing energy density that is required for the warp bubble dynamics with respect to bulk space.

## 6. Discussion

Figure 5 shows that the warp bubble is pushed by the underlying bulk space of the braneworld model. The expression of brane velocity has been developed by analysing the property of the hyperspace metric by obtaining the extremum expression of proper time from equation (4). By finding the Lagrangian of the equations (9) and (10), we derived the warp bubble velocity where proper time is exactly the same with relativistic time, which indicates that the velocity is not constrained by the speed of light. This can also prove timelike characteristic of the warp bubble interior, thus ensuring that the vehicle inside the bubble remains stationary, whereas the bubble itself may move at a velocity with no special relativistic constrain. This leads to the notion that the inertial reference frame is preserved inside warp bubble. A more generalized metric tensor has been derived as shown in equation (22) which is similar to the Alqubierre spacetime metric expression. But for this case, the extradimensional term that represents bulk space is ready to be embedded for the calculation of the warp bubble extrinsic curvature after the shape function is properly defined as shown in Figure 2.

The shape function is defined by using tangent hyperbolic expression that neatly separate 4 regions with respect to the warp bubble radii as depicted in Figure 2 where it represents analogically as spacetime mould that is ready for more detail analysis using warp bubble brane velocity terms derived from the extrinsic curvature expression and expressed in term of dynamics characteristic (38). Figure 3 shows the extrinsic curvature of the warp bubble hypersurface interacting with the surrounding space.

The Lie derivative of the projection tensor of equation (28) represents the extrinsic curvature which can be expanded as a 3 by 3 matrix. Considering the warp bubble as perfect spherical symmetric, the trace of the matrix represents the dynamics characteristic that indicates the warp bubble surroundings either expanding or contracting which is later used to describe the warp bubble shape function characteristics in greater detail. Since the bulk extradimensional term $Y(r)$ is included in the dynamic characteristic expression (34), the characteristic expression can be expressed in terms of bulk and brane velocities (38). The sum of these two velocities is the hyperdrive velocity $v_{HD}$. The shape function based on the tangent hyperbolic function at 3 different radii conditions of equations (25), (26) and (27) was analysed by using the dynamic characteristic $\chi$ of equation (38). The result has shown that the slope of the transition region which is representing the bubble thickness $\xi$ has influence on the warp bubble geometrodynamics. These indicate that the warp bubble propulsion is by the geometrodynamics of contraction and expansion of hyperspace surrounding the warp bubble, thus leads to the notion that the motion inertial reference frame may preserved in bubble's interior region so that the reference frame in the bubble's interior and the exterior is completely separated.

Finally, by the energy momentum tensor (42) derived from Einstein curvature tensor, it has been shown that the energy density acquires exotic matter property of negative energy at the slope of the transition region. This indicates that beside the hyperdrive energy density $\rho_{H\_warp}$ (47), which consist of brane and bulk components (48), the thickness of the bubble's boundary $\xi$ also plays an important factor in determining the velocity of the warp bubble, thus the warp bubble dynamics is not constrained by special relativity limitation. As the energy density has the brane and bulk components, the velocity of the warp bubble also consists of the brane and bulk components and the figure 5 shows figuratively *kite surfing-like dynamics* of embedded hypersurface onto a hyperspace of braneworld. This shows the underlying extra-dimensional of hyperspace influence on the warp bubble geometrodynamics on the brane which is the 3-dimensional space universe. In another words, the braneworld "floats" in 4 spatial dimensional bulk of hyperspace.

## 7. Conclusion

The velocity of a hyperdrive warp bubble is as superposition of the velocity of the warp bubble on the brane and the velocity of the underlying bulk. Therefore, the hyperdrive velocity is larger than the warp drive velocity that is occurring only on the brane whereas the hyperdrive velocity is the addition of the underlying bulk velocity to the warp drive velocity on the brane. Similar mathematical characteristics of superposition also occur for the hyperdrive energy density which resulted from the addition of brane energy density and bulk energy density. Therefore, the hyperdrive requires much more energy than that of a warp drive. A hyperdrive that utilizes the underlying higher dimensional spatial energy of bulk space is analogous to a *kite surfer* that uses *wind* as the bulk force, which adds up to the dynamics of surfing, which is originally propelled by the *surfing wave motion* analogous to the brane.

## 8. Acknowledgement

The authors would like to thank Universiti Kebangsaan Malaysia for funding the work of this research under grant GGPM-2022-038.